\documentclass[%
 preprint,
showpacs,
nofootinbib,
 amsmath,amssymb,
 prc,
 superscriptaddress
]{revtex4-1}

\usepackage{graphicx}
\usepackage{dcolumn}
\usepackage{bm}
\usepackage{braket}
\usepackage{caption}
\usepackage{subcaption}

\begin{document}


\title{Repulsive Four-Body Interactions of $\alpha$ Particles and Quasi-Stable Nuclear $\alpha$-Particle Condensates in Heavy Self-Conjugate Nuclei}


\author{Dong Bai}
 \email{dbai@itp.ac.cn}
\affiliation{School of Physics, Nanjing University, Nanjing, 210093, China}%

\author{Zhongzhou Ren}
\email{Corresponding Author: zren@tongji.edu.cn}
\affiliation{School of Physics Science and Engineering, Tongji University, Shanghai, 200092, China}%


\date{\today}

\begin{abstract}
We study the effects of repulsive four-body interactions of $\alpha$ particles on nuclear $\alpha$-particle condensates in heavy self-conjugate nuclei using a semi-analytic approach, and find that the repulsive four-body interactions could decrease the critical number of $\alpha$ particles, beyond which quasi-stable $\alpha$-particle condensate states can no longer exist, even if these four-body interactions make only tiny contributions to the total energy of the Hoyle-like state of $^{16}$O. Explicitly, we study eight benchmark parameter sets, and find that the critical number $N_\text{cr}$ decreases by $|\Delta N_\text{cr}|\sim1-4$ from $N_\text{cr}\sim11$ with vanishing four-body interactions. We also discuss the effects of four-body interactions on energies and radii of $\alpha$-particle condensates. Our study can be useful for future experiments to study $\alpha$-particle condensates in heavy self-conjugate nuclei. Also, the experimental determination of $N_\text{cr}$ will eventually help establish a better understanding on the $\alpha$-particle interactions, especially the four-body interactions. 
\end{abstract}

\maketitle


Clustering phenomena are ubiquitous in both infinite and finite nuclear systems, and many important results have been obtained. For instance, Ref.~\cite{Ropke:1998qs} points out that nuclear $\alpha$-particle condensates could appear in the low-density symmetric nuclear matter. Other interesting works on clustering phenomena in symmetric and asymmetric nuclear matter include, e.g., Ref.~\cite{Takemoto:2004fp}, whose study is based on an impressive usage of density-fluctuated states. Nuclear $\alpha$-particle condensates are conjectured to play an important role also in describing Hoyle and Hoyle-like states in self-conjugate nuclei such as $^{12}$C, $^{16}$O, $^{20}$Ne, etc \cite{Tohsaki:2017hen}. It was proposed in 2001 that these states could be interpreted as gases of loosely bound $\alpha$ particles moving in the lowest $0$S orbit of their common mean field, similar to Bose-Einstein condensates observed in cold atomic systems \cite{Tohsaki:2001an}. Typically, these $\alpha$-particle condensate states, if exist, are excited $0^+$ states living energetically close to multi-$\alpha$ disintegration threshold, and occupy spatial volumes three or four times larger than the corresponding ground states. Many microscopic calculations have been carried out to study $\alpha$-particle condensates in light self-conjugate nuclei ($^{8}$Be, $^{12}$C, and $^{16}$O) theoretically, based on the generator coordinate method \cite{Uegaki:1977,Uegaki:1978}, orthogonality condition method (OCM) \cite{Horiuchi:1974,Horiuchi:1975,Yamada:2005ww,Funaki:2008gb}, antisymmetrized molecular dynamics \cite{Kanada-Enyo:2006rjf}, Tohsaki-Horiuchi-Schuck-R\"opke (THSR) wave function \cite{Tohsaki:2001an,Funaki:2002fn,Funaki:2014tda}, nuclear lattice effective field theory \cite{Epelbaum:2011md,Epelbaum:2013paa}, and Hartree-Fock-Bogoliubov approach \cite{Girod:2013faa}, etc. Furthermore, it is pointed out in Ref.~\cite{Funaki:2008rt} that the $\alpha$-particle condensate fraction is reduced as the nuclear density increases. Studies of nuclear $\alpha$-particle condensates also shed light on the criterion for Bose-Einstein condensation in traps and self-bound systems \cite{Yamada:2008}. We would like to recommend Ref.~\cite{Horiuchi:2012,Freer:2017gip} for comprehensive reviews on the field. It is now believed that there is no real counter argument that invalidates the picture of $\alpha$-particle condensates for, at least, the Hoyle state \cite{Tohsaki:2017hen}. On the other hand, situations in heavier nuclei are less studied both theoretically and experimentally. With the help of the Gross-Pitaevskii equation and Hill-Wheeler equation, Ref.~\cite{Yamada:2003cz} studies the $\alpha$-particle condensates in heavy self-conjugate nuclei, and conjectures that their quasi-stability could be sustained up to around $^{40}$Ca. To obtain this result, Ref.~\cite{Yamada:2003cz} considers two-body interactions between $\alpha$ particles, as well as \emph{repulsive} three-body interactions (or repulsive density-dependent potentials that play a similar role).
As a bonus of repulsive three-body interactions, the $\alpha$-particle condensates are stopped from falling gradually into collapsed states as the number of $\alpha$ particles increases.

In this note, we would like to study the impacts of non-vanishing repulsive four-body interactions of $\alpha$ particles on heavy $\alpha$-particle condensates. These four-body interactions could be resulted from nontrivial net effects of microscopic nuclear interactions and Pauli blocking. It is an important task to seek for a quantitative understanding of four-body interactions of $\alpha$ particles from the microscopic viewpoint, which is quite challenging and lies beyond the scope of the present work. As a compromise, we study four-body interactions from the pure phenomenological viewpoint, inspired by the fact that they are crucial to reproduce the spectrum of $^{16}$O as revealed by the explicit OCM calculation \cite{Funaki:2008gb}. This is also the viewpoint taken by Yamada and Schuck in Ref.~\cite{Yamada:2003cz} when they introduce three-body interactions of $\alpha$ particles. In general, it is very difficult to solve nuclear many-body problems from the first principle. Often, one switches to some effective Hamiltonians with effective interactions introduced to reproduce several essential physical properties. The two-body effective interactions between $\alpha$ particles, such as the Ali-Bodmer interactions \cite{Ali:1966}, are widely used in literature to study nuclear structures and reactions of $N\alpha$ systems. Further studies show that there could also be three-body interactions of $\alpha$ particles, which are crucial for reproducing the spectrum of $^{12}$C \cite{Fukatsu:1992}. It is quite natural to expect that non-vanishing four-body interactions of $\alpha$ particles also play an important role. Indeed, the explicit OCM calculation shows that repulsive four-body interactions are needed to fit the ground-state energy of $^{16}$O \cite{Funaki:2008gb}. Repulsive four-body interactions, rather than attractive ones, could also help avoid the appearance of collapsed states in the $N\alpha$ system, similar to what repulsive three-body interactions do as discussed in Ref.~\cite{Yamada:2003cz}.


Following Ref.~\cite{Yamada:2003cz}, we treat $\alpha$ particles as structureless bosons, and adopt the mean-field viewpoint by depicting $\alpha$-particle condensate states as products of identical single-particle wave functions
\begin{align}
\Psi(N\alpha)=\prod_{i=1}^N\varphi(\mathbf{r}_i),
\end{align}
where $\varphi$ is the normalized single-$\alpha$ wave function, and $\mathbf{r}_i$ is the coordinate of the $i$th $\alpha$ particle. This pure bosonic approach presumes the existence of tightly bound $\alpha$ particles (clusters) inside low-density nuclear $\alpha$-particle condensates and ignores their inner structures for simplicity. Unlike microscopic approaches which start typically with nucleon-nucleon interactions and treat antisymmetrization effects exactly, the pure bosonic approach adopts effective interactions of $\alpha$ particles, with antisymmetrization effects simulated entirely by repulsive interactions at the short distance. Compared with microscopic approaches, the pure bosonic approach certainly cannot be viewed as fundamental and might not be applicable for nuclear systems with higher densities where $\alpha$ particles dissolve eventually due to the Pauli blocking. Also, $\alpha$-particle condensate fraction is reduced as the nuclear density increases \cite{Funaki:2008rt}. The pure bosonic approach is however suitable for our current purposes to study low-density nuclear $\alpha$-particle condensates as shown in previous studies \cite{Yamada:2003cz,Funaki:2008rt,Yamada:2008}. The total energy of the $N\alpha$ system is then given by
\begin{align}
&E(N\alpha)\nonumber\\
&\!\!\!\!\!\!\!\!=\braket{\Psi(N\alpha)|H|\Psi(N\alpha)}\nonumber\\
&\!\!\!\!\!\!\!\!=N\bigg[\braket{T_N}+\frac{1}{2}(N-1)\braket{V_2}+\frac{1}{6}(N-1)(N-2)\braket{V_3}+\frac{1}{24}(N-1)(N-2)(N-3)\braket{V_4}
\bigg],
\end{align}
with
\begin{align}
&\quad\quad\quad\quad\,
\braket{T_N}=\left(1-{1}/{N}\right)\braket{\varphi(\mathbf{r})|-\frac{\hbar^2}{2m}\nabla^2\,|\varphi(\mathbf{r})},
\\
&\quad\quad\quad\quad
\braket{V_2}=\braket{\varphi(\mathbf{r}_1)\varphi(\mathbf{r}_2)|V_2(\mathbf{r}_1,\mathbf{r}_2)|\varphi(\mathbf{r}_1)\varphi(\mathbf{r}_2)},\label{V2}
\\
&\quad\quad
\braket{V_3}=\braket{\varphi(\mathbf{r}_1)\varphi(\mathbf{r}_2)\varphi(\mathbf{r}_3)|V_3(\mathbf{r}_1,\mathbf{r}_2,\mathbf{r}_3)|\varphi(\mathbf{r}_1)\varphi(\mathbf{r}_2)\varphi(\mathbf{r}_3)},
\\
&
\braket{V_4}=\braket{\varphi(\mathbf{r}_1)\varphi(\mathbf{r}_2)\varphi(\mathbf{r}_3)\varphi(\mathbf{r}_4)|V_4(\mathbf{r}_1,\mathbf{r}_2,\mathbf{r}_3,\mathbf{r}_4)|\varphi(\mathbf{r}_1)\varphi(\mathbf{r}_2)\varphi(\mathbf{r}_3)\varphi(\mathbf{r}_4)},
\end{align}
as the single-$\alpha$ kinetic energy with the center-of-mass correction, the two-body potential energy, the three-body potential energy, and the four-body potential energy, respectively. In general, the energy functional $E(N\alpha)$ is a nonlocal functional of the wave function $\varphi(\mathbf{r})$.

We adopt the Ali-Bodmer ansatz for nuclear interactions \cite{Ali:1966}:
\begin{align}
&V_\text{2N}(\mathbf{r}_1,\mathbf{r}_2)=V_r\exp(-\mu_r^2|\mathbf{r}_1-\mathbf{r}_2|^2)-V_a\exp(-\mu_a^2|\mathbf{r}_1-\mathbf{r}_2|^2),\label{ABP}\\
&V_3(\mathbf{r_1},\mathbf{r}_2,\mathbf{r}_3)=W_3\exp\{-\mu_\omega^2[(\mathbf{r}_1-\mathbf{r}_2)^2+(\mathbf{r}_1-\mathbf{r}_3)^2+(\mathbf{r}_2-\mathbf{r}_3)^2]\},\\
&V_4(\mathbf{r}_1,\mathbf{r}_2,\mathbf{r}_3,\mathbf{r}_4)=U_4\exp\{-\mu_\upsilon^2[(\mathbf{r}_1-\mathbf{r}_2)^2+(\mathbf{r}_1-\mathbf{r}_3)^2+(\mathbf{r}_1-\mathbf{r}_4)^2\nonumber\\
&\qquad+(\mathbf{r}_2-\mathbf{r}_3)^2+(\mathbf{r}_2-\mathbf{r}_4)^2+(\mathbf{r}_3-\mathbf{r}_4)^2]\}.
\end{align}
For the Coulomb interaction, we take
\begin{align}
V_\text{2C}(\mathbf{r}_1,\mathbf{r}_2)=\frac{4e^2}{|\mathbf{r}_1- \mathbf{r}_2|}\text{erf}\!\left(\frac{\sqrt{3}}{2R_\alpha}\left|\mathbf{r}_1-\mathbf{r}_2\right|\right),\quad\quad R_\alpha=1.44\text{ fm},
\end{align}
as inspired by double-folding potential calculations \cite{Ali:1966,Ren:1994zz}.

\begin{figure}[tb]
\centering
\includegraphics[width=0.9\textwidth]{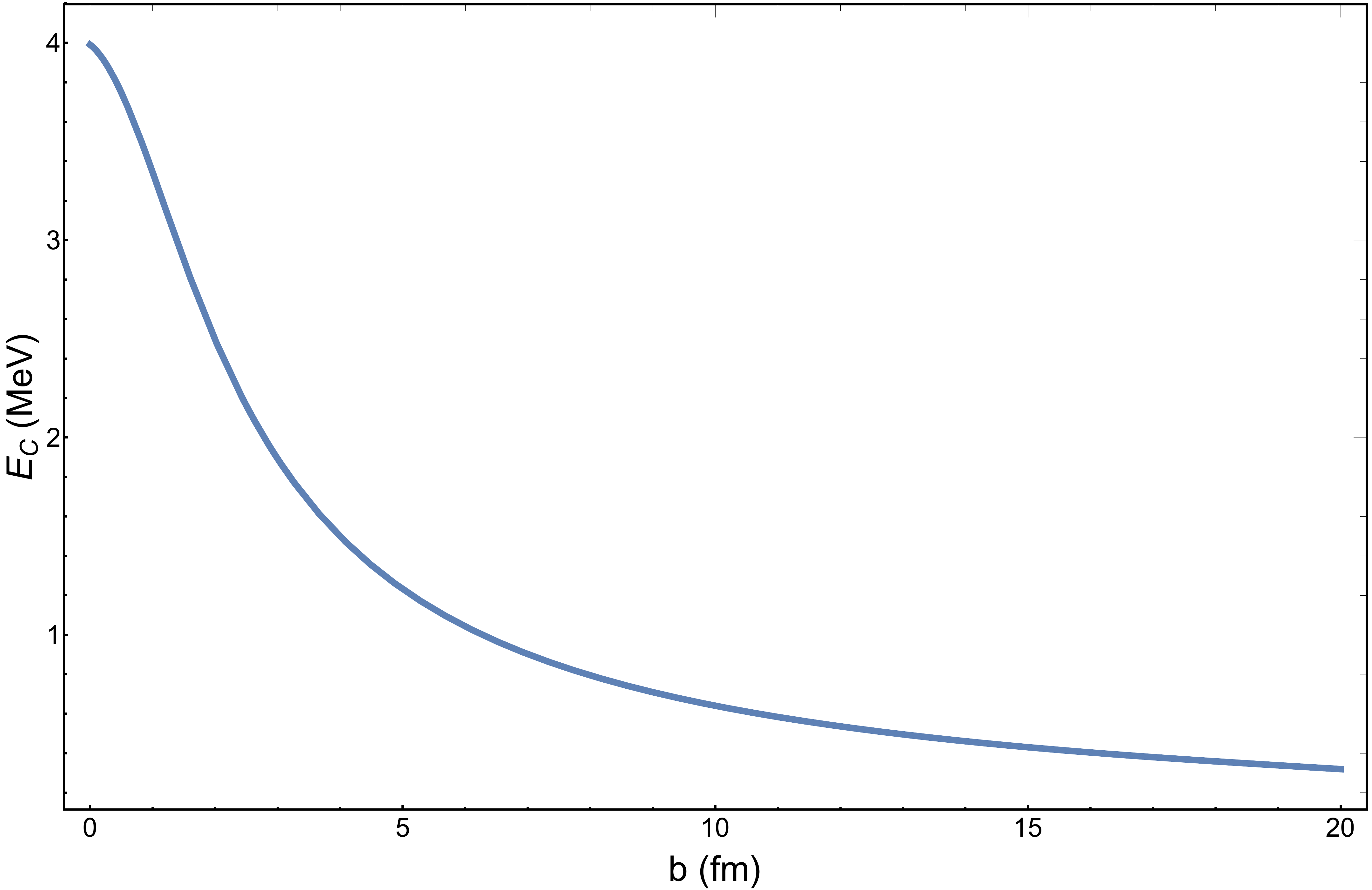}
\caption{Coulomb functional $\mathcal{E}_\text{C}(b)$ for $b\in [0,20]$ fm.}
\label{CE} 
\end{figure}

With the above ansatz, we can estimate the energy functional $E(N\alpha)$ easily by using the harmonic-oscillator wave function
\begin{align}
\varphi_b(\mathbf{r})=\left(\!\frac{2}{\pi b^{2}}\!\right)^{\!\!3/4}\!\!\!\!\exp\!\left(-{r^2}/{b^2}\right).
\end{align}
With the Coulomb energy function (plotted in Fig.~\ref{CE})
\begin{align}
\mathcal{E}_\text{C}(b)=\!\int\!\mathrm{d}^3r_1\mathrm{d}^3r_2|\varphi_b(\mathbf{r}_1)|^2|\varphi_b(\mathbf{r}_2)|^2\frac{4e^2}{|\mathbf{r}_1-\mathbf{r}_2|}\text{erf}\!\left(\frac{\sqrt{3}}{2R_\alpha}\left|\mathbf{r}_1-\mathbf{r}_2\right|\right),
\end{align}
the energy functional $E(N\alpha)$ could be given semi-analytically by
\begin{align}
E(N\alpha)&=(N-1)\frac{3\hbar^2}{2b^2m}+\frac{1}{2}N(N-1)\!\left[\frac{V_r}{(b^2\mu_r^2+1)^{3/2}}-\frac{V_a}{(b^2\mu_a^2+1)^{3/2}}+\mathcal{E}_\text{C}(b)\right]\nonumber\\
&+\frac{1}{6}N(N-1)(N-2)\frac{8W_3}{(3b^2\mu_\omega^2+2)^3}\nonumber\\
&+\frac{1}{24}N(N-1)(N-2)(N-3)\frac{U_4}{(2b^2\mu_\upsilon^2+1)^{9/2}}.\label{ENaAB}
\end{align}
The harmonic-oscillator parameter $b$ is related to the charge root-mean-square (rms) radius of the $N\alpha$ state by
\begin{align}
&\sqrt{\braket{r^2_N}}=\sqrt{\braket{r^2_\alpha}_\text{GP}+1.71^2},\\
&\braket{r^2_\alpha}_\text{GP}=\left(1-1/N\right)\braket{\varphi_b|r^2|\varphi_b}=(1-1/N)\frac{3}{4}b^2.
\end{align}
Here, we have considered explicitly the finite-size effect of the $\alpha$ particle and the center-of-mass correction. With these formulae, for a given $N$ we can obtain the energies and radii of the $\alpha$-particle condensate state directly by minimizing $E(N\alpha)$ with respect to $b$.

Before investigating the physical effects of four-body interactions, let's first check the robustness of our method by redoing the analysis in Ref.~\cite{Yamada:2003cz} using the same set of parameters. Explicitly, we take the Yamada-Schuck parameter set
\begin{align}
&V_r=50 \text{ MeV}, \quad V_a=34.101 \text{ MeV}, \quad W_3=151.5 \text{ MeV},\nonumber\\
&\mu_r=0.4 \text{ fm}^{-1}, \quad \mu_a=0.3 \text{ fm}^{-1}, \quad \mu_\omega=0.387 \text{ fm}^{-1},\label{YSPotential}
\end{align}
without four-body interactions. The parameters of the two-body interaction are chosen to reproduce the ground-state energy of $^{8}$Be, with the radial part of the wave function being small in the inner region and having a maximum value around $r=4$ fm, while the parameters of the three-body interaction are adopted from previous OCM studies on $^{12}$C \cite{Fukatsu:1992}. Noticeably, the two-body interactions depicted by Eq.~\eqref{YSPotential} are softer than the famous Ali-Bodmer two-body potentials \cite{Ali:1966} which is found to be not suitable for the studies of $\alpha$-particle condensate states \cite{Yamada:2003cz}. The numerical results are listed in Table \ref{3aForceResults}, along with results obtained by using the Gross-Pitaevskii equation in Ref.~\cite{Yamada:2003cz} for comparison. It is straightforward to see that, the energies and radii obtained from the semi-analytic energy function Eq.~\eqref{ENaAB} agree quite well with those in Ref.~\cite{Yamada:2003cz}, with tiny differences to be roughly 1\% and 8\% for the energies and radii, respectively. This could be viewed as an impressive test for the robustness of our semi-analytic approach. Compared with solving the Gross-Pitaevskii equation numerically, its main advantage is that the computational cost is low and the accuracy loss is tiny. Furthermore, similar to Ref.~\cite{Yamada:2003cz} we also observe the decreasing behaviors of the Coulomb barriers with respect to increasing $N$ values. With the Yamada-Schuck model Eq.~\eqref{YSPotential}, the critical number of $\alpha$ particles that the $\alpha$-particle condensate state can sustain to be quasi-stable is estimated to be $N_\text{cr}=11$, consistent with Ref.~\cite{Yamada:2003cz}. 

\begin{table}
\caption{Energies and rms radii of $N\alpha$ states with the parameters for two-body and three-body interactions given by the Yamada-Schuck parameter set Eq.~\eqref{YSPotential}. All energies are given in the unit of MeV, and all lengths are given in the unit of fm. Numerical results from the Gross-Pitaevskii equation in Ref.~\cite{Yamada:2003cz} have the additional subscript ``YS''. }
\label{3aForceResults}
\begin{center}
\begin{tabular}{ccccccccc}
\hline
\hline
\hspace{5mm}$N$\hspace{8mm} & \hspace{5mm}{Nucleus}\hspace{5mm} 
     & \hspace{5mm}$E$\hspace{5mm} & \hspace{5mm}$\sqrt{\braket{r^2_N}}$\hspace{5mm}
     & \hspace{5mm}$E_\text{YS}$\hspace{5mm} & \hspace{5mm}$\sqrt{\braket{r^2_N}}_\text{YS}$\hspace{5mm} \\[0.5ex]  
\hline
 3 & $^{12}$C   & 0.99 & 4.88 & 0.98 & 4.87 \\
 4 & $^{16}$O   & 1.86 & 5.34 & 1.84 & 5.23 \\
 5 & $^{20}$Ne & 3.09 & 5.74 & 3.04 & 5.55 \\
 6 & $^{24}$Mg & 4.70 & 6.10 & 4.63 & 5.85 \\
 7 & $^{28}$Si   & 6.70 & 6.44 & 6.61 & 6.13 \\
 8 & $^{32}$S    & 9.10 & 6.77 & 8.99 & 6.40 \\
 9 & $^{36}$Ar   & 11.88 & 7.10 & 11.8 & 6.68 \\
10 & $^{40}$Ca & 15.05 & 7.46 & 15.0 & 6.95 \\
11 & $^{44}$Ti   & 18.58 & 7.86 & 18.6 & 7.24 \\
\hline
\hline
\end{tabular}
\end{center}
\end{table}

In the rest part of this note, we shall adopt the semi-analytic approach to investigate the effects of four-body interactions on the heavy $\alpha$-particle condensate systems. As revealed in Eq.~\eqref{ENaAB}, the four-body potential energy scales as $\braket{V_4}\sim N(N-1)(N-2)(N-3)/24$, which grows faster than the kinetic energy $\braket{T_N}\sim N-1$, the two-body potential energy $\braket{V_2}\sim N(N-1)/2$, and the three-body potential energy $\braket{V_3}\sim N(N-1)(N-2)/6$ as $N$ increases. Due to the complexity of nuclear many-body problems, estimating the parameters of four-body interactions from the first principle is not an easy job and lies beyond the scope of our present study. Instead, in this note, we shall consider eight different benchmark parameter sets for four-body interactions and try to study their physical effects separately. 
These eight benchmark parameter sets are given as follows:
\begin{itemize}
\item Parameter set $\vartriangleleft\!4\alpha\!\vartriangleright$:
\begin{align}
U_4 = 2500 \text{ MeV},\quad\quad\quad\quad \mu_\upsilon= 0.387 \text{ fm}^{-1}; \label{SR}
\end{align}
\item Parameter set $\diamondsuit\,4\alpha\,\diamondsuit$:
\begin{align}
U_4 = 5000 \text{ MeV},\quad\quad\quad\quad \mu_\upsilon= 0.387 \text{ fm}^{-1};
\end{align}
\item Parameter set $\square\,4\alpha\,\square$:
\begin{align}
U_4 = 7500 \text{ MeV},\quad\quad\quad\quad \mu_\upsilon = 0.387 \text{ fm}^{-1};
\end{align}
\item Parameter set $\heartsuit\,4\alpha\,\heartsuit$:
\begin{align}
U_4 = 10000 \text{ MeV},\quad\quad\quad\quad \mu_\upsilon = 0.387 \text{ fm}^{-1};
\end{align}
\item Parameter set $\blacktriangleleft4\alpha\blacktriangleright$:
\begin{align}
U_4= 25 \text{ MeV},\quad\quad\quad\quad \mu_\upsilon= 0.2 \text{ fm}^{-1};
\end{align}
\item Parameter set $\blacklozenge\,4\alpha\,\blacklozenge$:
\begin{align}
U_4= 50 \text{ MeV},\quad\quad\quad\quad \mu_\upsilon= 0.2 \text{ fm}^{-1};
\end{align}
\item Parameter set $\blacksquare\,4\alpha\,\blacksquare$:
\begin{align}
U_4= 75 \text{ MeV},\quad\quad\quad\quad \mu_\upsilon= 0.2 \text{ fm}^{-1};
\end{align}
\item Parameter set $\spadesuit\,4\alpha\,\spadesuit$:
\begin{align}
U_4= 100 \text{ MeV},\quad\quad\quad\quad \mu_\upsilon= 0.2 \text{ fm}^{-1}. \label{LR}
\end{align}
\end{itemize}
Here, we consider two possibilities for the four-body interaction ranges. The two-body and three-body interactions are inherited from the Yamada-Schuck parameter set directly. 

\begin{table}
\caption{Energies and rms radii of various $\alpha$-particle condensate states for different benchmark parameter sets of four-body interactions as listed in Eq.~\eqref{SR}-\eqref{LR}. All energies and lengths are in the units of MeV and fm, respectively. The blanks refer to the fact that no quasi-stable $\alpha$-particle condensate states could be found for the target nucleus within the given interaction model.}
\label{4aResults}
\begin{center}
\begin{tabular}{cccccccccccccc}
\hline
\hline
      &
      & \multicolumn{2}{c}{$\vartriangleleft\!4\alpha\!\vartriangleright$} 
      & \multicolumn{2}{c}{$\diamondsuit\,4\alpha\,\diamondsuit$}
      & \multicolumn{2}{c}{$\square\,4\alpha\,\square$}
      & \multicolumn{2}{c}{$\heartsuit\,4\alpha\,\heartsuit$} \\
\hspace{4mm}$N$\hspace{4mm}  
      & \hspace{4mm}Nucleus\hspace{4mm}
      & \hspace{4mm}$E$\hspace{4mm} 
      & \hspace{4mm}$\sqrt{\langle r^2_N\rangle}$\hspace{4mm}
      & \hspace{4mm}$E$\hspace{4mm} 
      & \hspace{4mm}$\sqrt{\langle r^2_N\rangle}$\hspace{4mm}
      & \hspace{4mm}$E$\hspace{4mm} 
      & \hspace{4mm}$\sqrt{\langle r^2_N\rangle}$\hspace{4mm}
      & \hspace{4mm}$E$\hspace{4mm} 
      & \hspace{4mm}$\sqrt{\langle r^2_N\rangle}$\hspace{4mm}
      \\[0.5ex]
\hline
 4  &  $^{16}$O    & 1.88 & 5.38 & 1.89 & 5.42 & 1.90 & 5.45 & 1.91 & 5.49 \\
 5  &  $^{20}$Ne  & 3.13 & 5.83 & 3.17 & 5.91 & 3.21 & 5.99 & 3.24 & 6.06 \\
 6  &  $^{24}$Mg  & 4.78 & 6.25 & 4.85 & 6.38 & 4.91 & 6.50 & 4.96 & 6.61 \\
 7  &  $^{28}$Si    & 6.83 & 6.65 & 6.93 & 6.83 & 7.01 & 7.00 & 7.08 & 7.15 \\
 8  &  $^{32}$S     & 9.27 & 7.05 & 9.40 & 7.30 & 9.49 & 7.52 & 9.56 & 7.75 \\
 9  &  $^{36}$Ar    & 12.09 & 7.47 & 12.23 & 7.81 & 12.32 & 8.20 & - & - \\
 10 &  $^{40}$Ca  & 15.27 & 7.96 & - & - & - & - & - & - \\
 11 &  $^{44}$Ti    & - & - & - & - & - & - & - & - \\
\hline
\hline
\\
\\
\end{tabular}
\begin{tabular}{cccccccccccccc}
\hline
\hline
      &
      & \multicolumn{2}{c}{$\blacktriangleleft4\alpha\blacktriangleright$} 
      & \multicolumn{2}{c}{$\blacklozenge\,4\alpha\,\blacklozenge$}
      & \multicolumn{2}{c}{$\blacksquare\,4\alpha\,\blacksquare$}
      & \multicolumn{2}{c}{$\spadesuit\,4\alpha\,\spadesuit$} \\
\hspace{4mm}$N$\hspace{4mm}  
      & \hspace{4mm}Nucleus\hspace{4mm}
      & \hspace{4mm}$E$\hspace{4mm} 
      & \hspace{4mm}$\sqrt{\langle r^2_N\rangle}$\hspace{4mm}
      & \hspace{4mm}$E$\hspace{4mm} 
      & \hspace{4mm}$\sqrt{\langle r^2_N\rangle}$\hspace{4mm}
      & \hspace{4mm}$E$\hspace{4mm} 
      & \hspace{4mm}$\sqrt{\langle r^2_N\rangle}$\hspace{4mm}
      & \hspace{4mm}$E$\hspace{4mm} 
      & \hspace{4mm}$\sqrt{\langle r^2_N\rangle}$\hspace{4mm}
      \\[0.5ex]
\hline
 4  &  $^{16}$O    & 1.89 & 5.40 & 1.91 & 5.46 & 1.93 & 5.51 & 1.95 & 5.56 \\
 5  &  $^{20}$Ne  & 3.17 & 5.89 & 3.24 & 6.03 & 3.29 & 6.15 & 3.34 & 6.28 \\
 6  &  $^{24}$Mg  & 4.86 & 6.36 & 4.98 & 6.59 & 5.07 & 6.81 & 5.14 & 7.02 \\
 7  &  $^{28}$Si    & 6.95 & 6.83 & 7.12 & 7.18 & 7.23 & 7.54 & 7.30 & 7.96\\
 8  &  $^{32}$S     & 9.44 & 7.32 & 9.62 & 7.89 & - & - & - & - \\
 9  &  $^{36}$Ar    & 12.28 & 7.91 & - & - & - & - & - & - \\
 10 &  $^{40}$Ca  & - & - & - & - & - & - & - & - \\
 11 &  $^{44}$Ti    & - & - & - & - & - & - & - & - \\
\hline
\hline
\end{tabular}

\end{center}
\end{table}

\begin{figure}[tb]
\centering
\includegraphics[width=\textwidth]{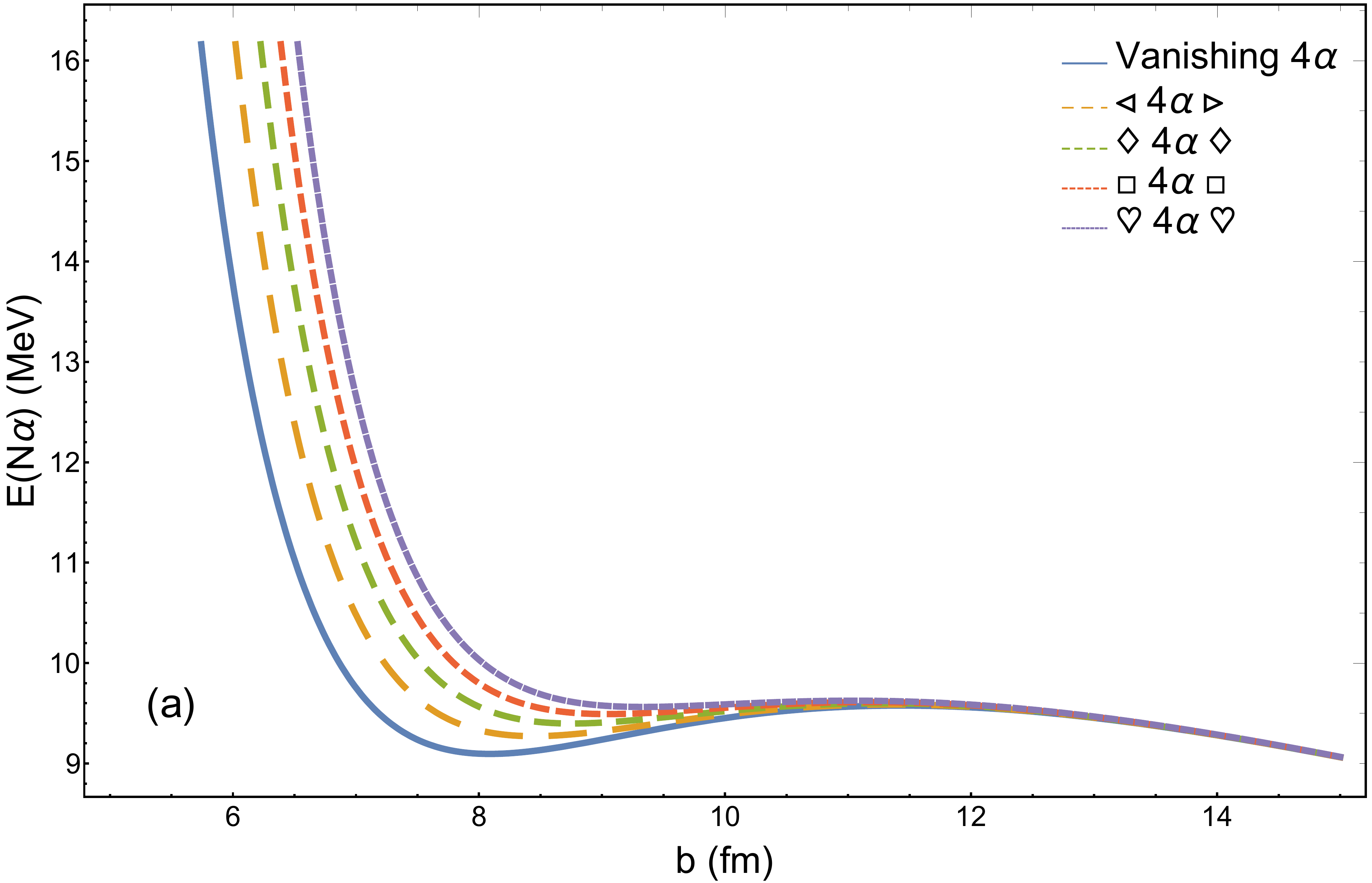}
\includegraphics[width=\textwidth]{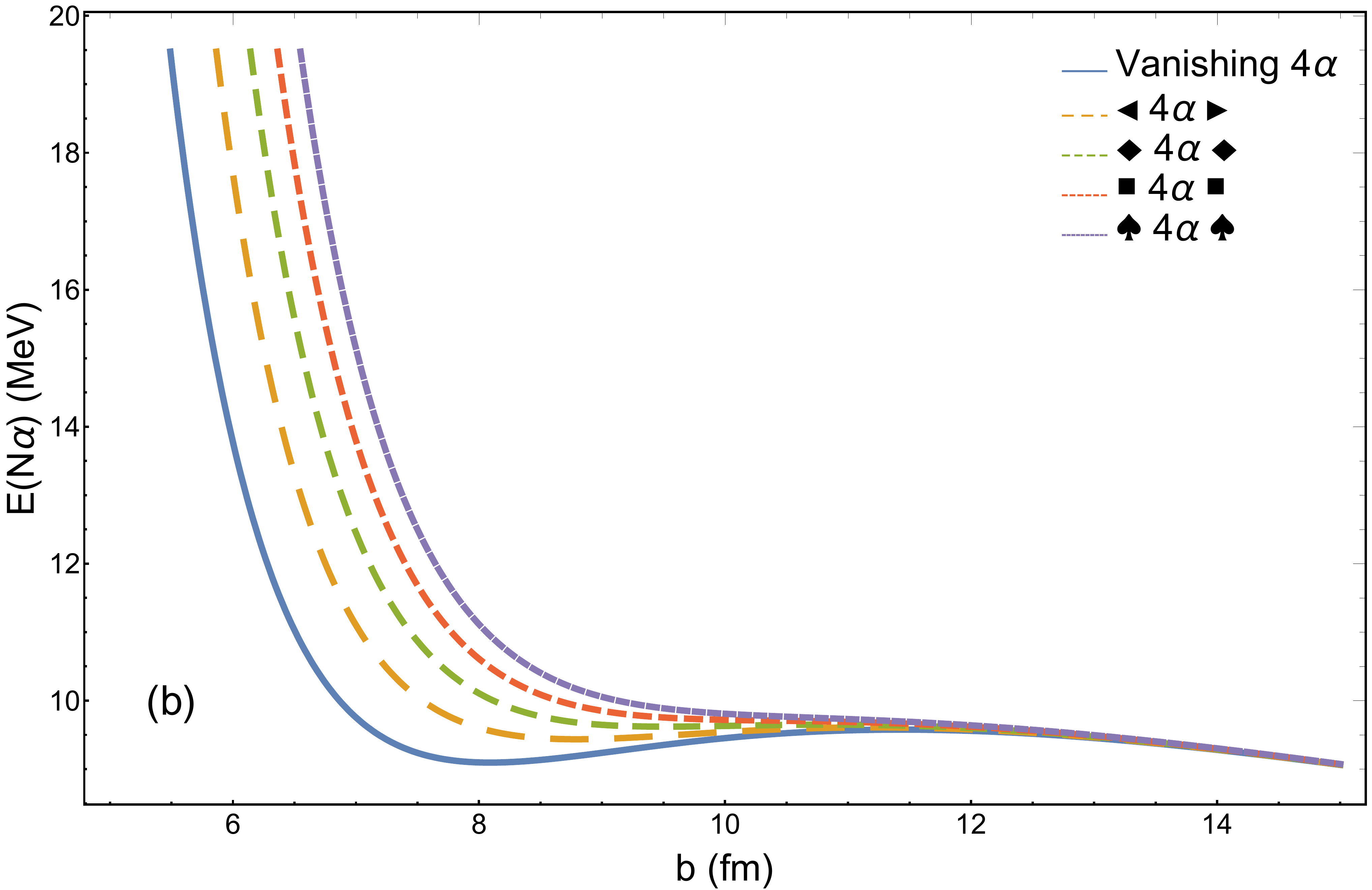}
\caption{Energy curves of $^{32}$S by different models of four-body interactions. Fig.~(a) and (b) display results for four-body interactions with $\mu_\upsilon=0.387 \text{ fm}^{-1}$ and $\mu_\upsilon=0.2 \text{ fm}^{-1}$, respectively.}
\label{S32} 
\end{figure}

The numerical results are presented in Table \ref{4aResults}. The four-body interactions are chosen carefully such that their contributions to the $^{16}$O total energy are tiny, less than 5\% for all benchmark parameter sets, much smaller than the three-body-force contributions being around 18\%. Given this fact, it might be a bit surprising to see that such ``tiny'' four-body interactions eventually play a crucial role in the $\alpha$-particle condensate formation for large $N$s. Here, the critical value that the $N\alpha$ system could sustain quasi-stable is determined by the vanishing of Coulomb barriers following Ref.~\cite{Yamada:2003cz}. Depending on which benchmark parameter sets is talked about, the critical value $N_\text{cr}$ gets a reduction $|\Delta N_\text{cr}|\sim1-4$, and gives $N_\text{cr}\sim7-10$. It is interesting to note that predictions from four-body interactions with different interaction ranges but roughly the same strength (i.e., making the same contribution to the $^{16}$O total energy) look similar to each other, which might point to some model-independent aspects of our analysis. Let's take $^{32}$S as an example to see more explicitly the effects of repulsive four-body interactions on Coulomb barriers. The corresponding energy curves could be found in Fig.~\ref{S32}. We can see that the repulsive four-body interactions generally decrease the heights of Coulomb barriers, and make the $N\alpha$ state less stable. The stronger the repulsive four-body interactions are, the shallower the energy well becomes and the smaller $N_\text{cr}$ is. 
  
\begin{figure}
\centering

\begin{subfigure}[b]{16cm}
\centering
\includegraphics[width=16cm]{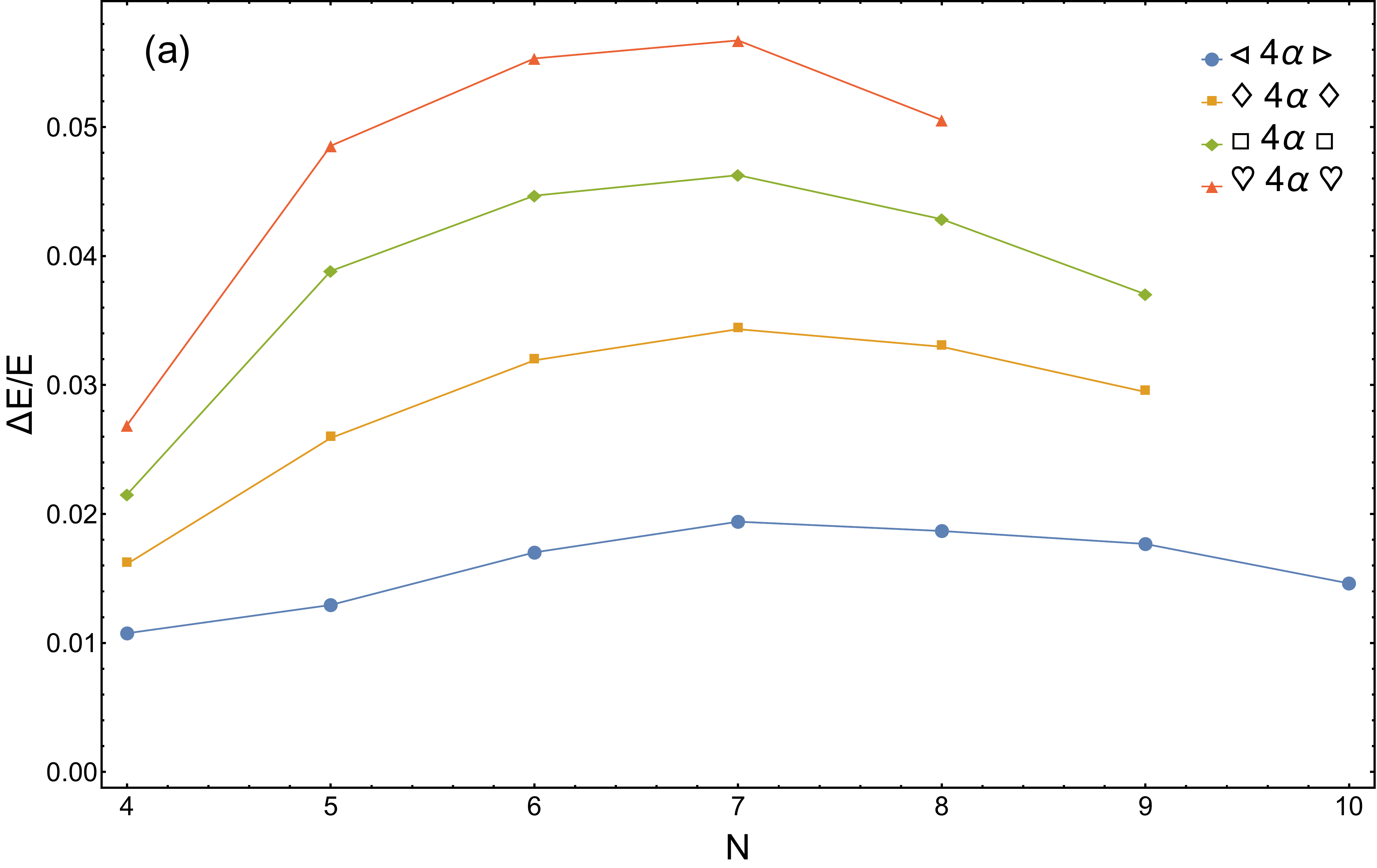}
\end{subfigure}

\begin{subfigure}[b]{16cm}
\centering
\includegraphics[width=16cm]{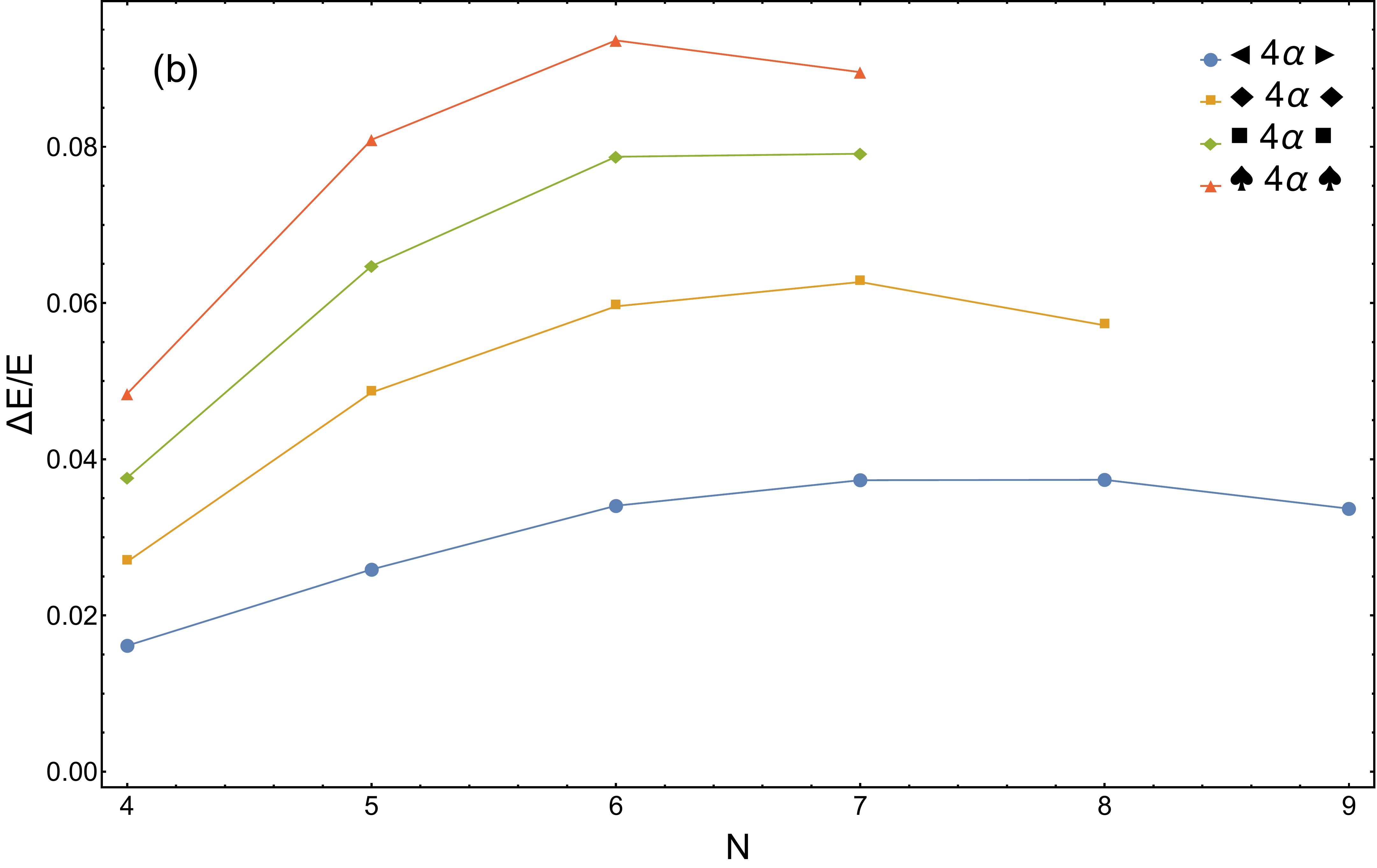}
\end{subfigure}

\caption{Relative changes in total energies of various $\alpha$-particle condensate states measured from the $N\alpha$ threshold by different benchmark parameter sets of four-body interactions. Fig.~(a) and (b) display results for four-body interactions with $\mu_\upsilon=0.387 \text{ fm}^{-1}$ and $\mu_\upsilon=0.2 \text{ fm}^{-1}$, respectively.}
\label{EVSANum}
\end{figure}

\begin{figure}
\centering

\begin{subfigure}[b]{16cm}
\centering
\includegraphics[width=16cm]{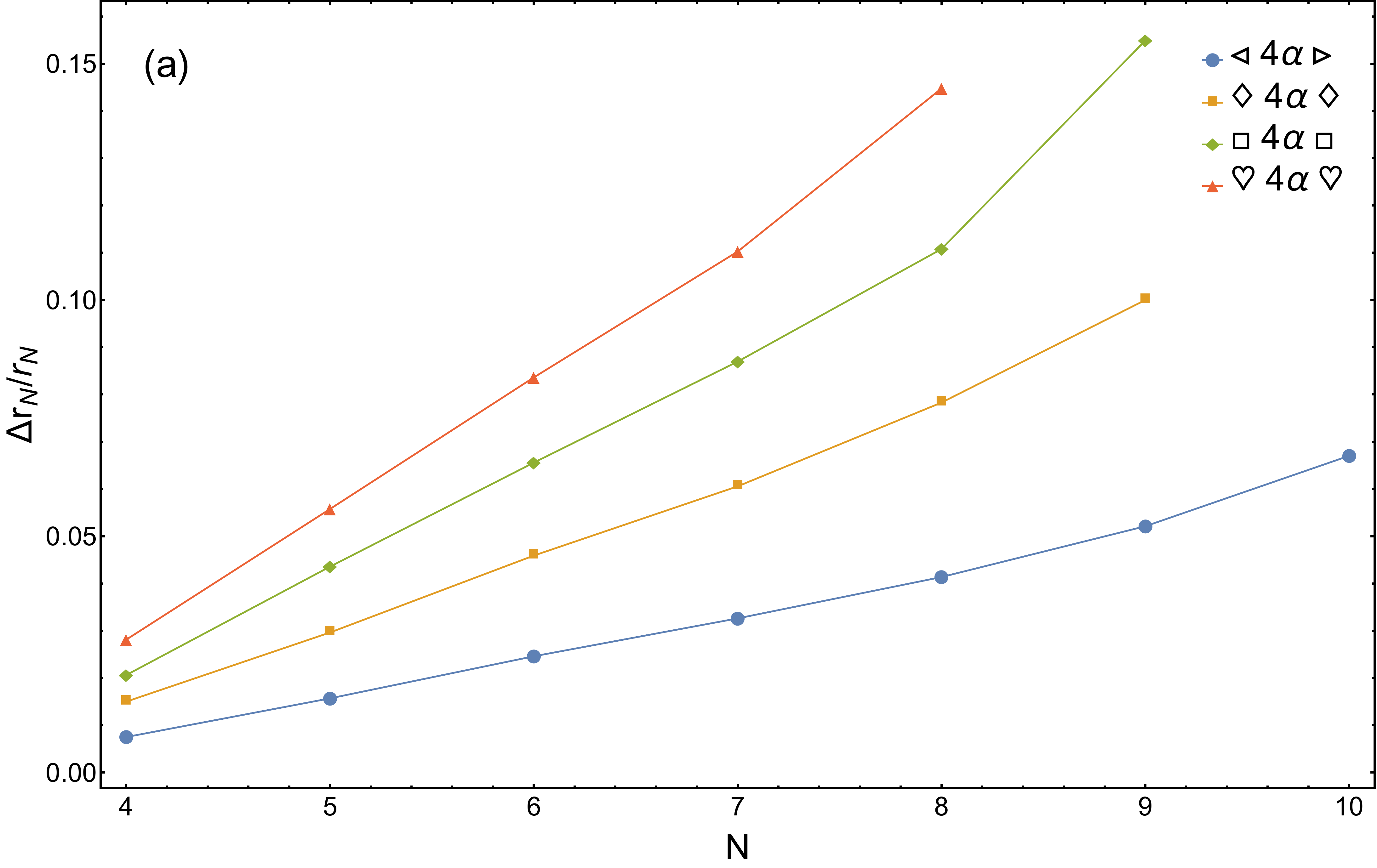}
\end{subfigure}

\begin{subfigure}[b]{16cm}
\centering
\includegraphics[width=16cm]{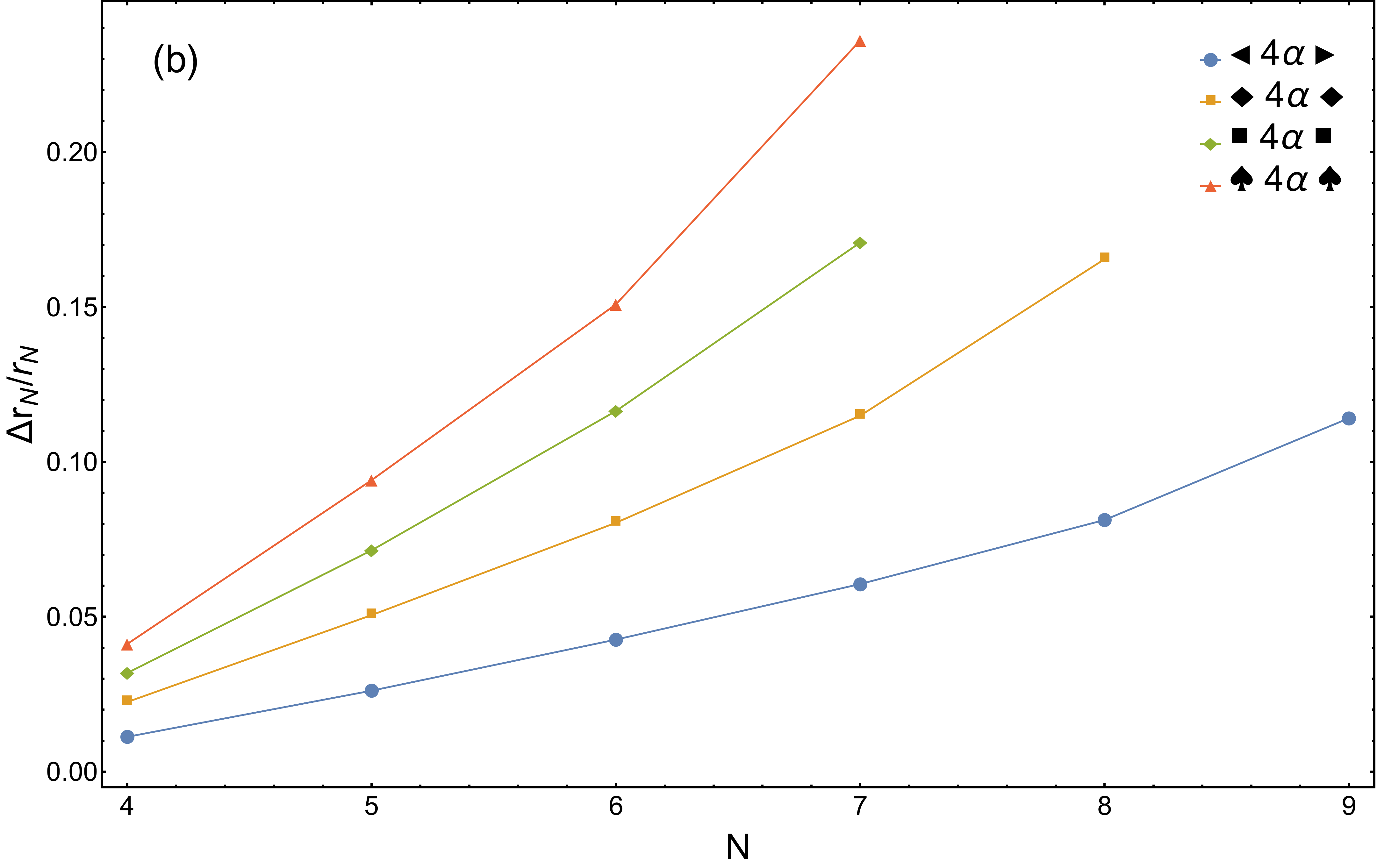}
\end{subfigure}

\caption{Relative changes in rms radii of various $\alpha$-particle condensate states measured from the $N\alpha$ threshold by different benchmark parameter sets of four-body interactions. Fig.~(a) and (b) display results for four-body interactions with $\mu_\upsilon=0.387 \text{ fm}^{-1}$ and $\mu_\upsilon=0.2 \text{ fm}^{-1}$, respectively.}
\label{RVSANum}
\end{figure}

Also, we find that, repulsive four-body interactions could cause small increases in total energies (less than 10\%) and rms radii (less than 30\%) of the $N\alpha$ systems. This is consistent with the physical intuition that repulsive interactions expel two particles away and contribute positively to the total energy. In Fig.~\ref{EVSANum} and \ref{RVSANum}, we plot the relative changes of total energies and radii for each benchmark parameter sets. It is interesting to note that there is a peak around $N=6\sim7$ in relative changes of total energies, while the relative changes of radii increase monotonically. Moreover, in Fig.~\ref{EVSANum} and \ref{RVSANum}, more hints might be found for the weak dependence on interaction ranges of our analysis. For instance, Model $\diamondsuit\,4\alpha\,\diamondsuit$ and Model $\blacktriangleleft4\alpha\blacktriangleright$, where the four-body interactions both contribute roughly 1.5\% to the $^{16}$O total energy, show similar behaviors in relative changes of both total energies and radii.

In summary, in this note we study the effects of four-body interactions of $\alpha$ particles on properties of nuclear $\alpha$-particle condensates in heavy self-conjugate nuclei. In our treatment, the four-body interactions are introduced from the phenomenological viewpoint as inspired by the explicit OCM calculations on $^{16}$O. This is also the viewpoint taken by Yamada and Schuck in Ref.~\cite{Yamada:2003cz} when they introduce three-body interactions of $\alpha$ particles. The question of the microscopic origin of $\alpha$-particle many-body interactions is important, insightful, and challenging, lying beyond the scope of the present work. Our semi-analytic framework is based on the mean-field theory similar to that of Ref.~\cite{Yamada:2003cz}, and could give accurate results with almost negligible computational costs. To account for theoretical uncertainties, we introduce eight benchmark parameter sets by incorporating the extra repulsive four-body interactions upon the Yamada-Schuck parameter set. The interaction strengths for the four-body interactions are chosen such that they make only tiny contributions to the total energy of the $\alpha$-particle condensate state in $^{16}$O, less than 5\% for all benchmark parameter sets. Such tiny four-body interactions are found to play an important role in determining the critical value $N_\text{cr}$ beyond which there are no longer any quasi-stable $\alpha$-particle condensate states. Explicitly, we have $N_\text{cr}\sim7-10$ for the benchmark parameter sets, smaller than $N_\text{cr}\sim11$ in the Yamada and Schuck's study. Besides, repulsive four-body interactions could also increase the total energies and radii of the $N\alpha$ systems. Our study also provides some hints for the possibility that the effects of repulsive four-body interactions depend mainly on their interaction strengths. In general, the stronger the four-body interactions are, the larger effects they cause. The analysis here could certainly be generalized further to five-body interactions. Although interesting, studies on five-body interactions go beyond the scope of the present work, and their effects will be viewed as theoretical uncertainties to be resolved in the future. Our work can be viewed as a useful complement to Ref.~\cite{Yamada:2003cz}, and could be important also for future experimental studies of quasi-stable $\alpha$-particle condensates in heavy self-conjugate nuclei. The experimental determination of the critical value $N_\text{cr}$, when coming true, will provide us with rich information on interactions between $\alpha$ particles, especially the four-body interactions. There have already been some attempts in pushing forward this frontier \cite{Akimune:2013,Swartz:2015lja,Bishop:2017csu}. Hopefully, we shall not wait too long to witness some progress.

\begin{acknowledgments} 
D.~B.~would like to thank Taiichi Yamada, Peter Schuck, and Yasuro Funaki for helpful communications. Also, we would like to thank the anonymous referee for his/her  suggestions and helps. This work is supported by the National Natural Science Foundation of China (Grant No.~11535004, 11761161001, 11375086, 11120101005, 11175085 and 11235001), by the National Major State Basic Research and Development of China, Grant No.~2016YFE0129300, and by the Science and Technology Development Fund of Macau under Grant No.~068/2011/A.
\end{acknowledgments}

\end{document}